\documentclass[aps,amssymb,amsmath,amsfonts,twocolumn,pra,showpacs]{revtex4}

\usepackage{amssymb,amsmath,amsthm,color,graphicx,times,bbold,graphicx}
\usepackage{hyperref}
\usepackage{enumerate}
\usepackage{subfigure}

\begin{document}

\title{Non-Markovianity of Gaussian Channels}


\author{G. Torre$^{1,2}$, W. Roga$^{3,4}$, and F. Illuminati$^{2,3}$\footnote{Corresponding author: illuminati@sa.infn.it}}
\affiliation{$^1$ Dipartimento di Fisica ``E. R. Caianiello'', Universit\`a degli Studi di Salerno, Via Giovanni Paolo II 132, I-84084 Fisciano (SA),Italy}
\affiliation{$^2$ INFN, Sezione di Napoli, Gruppo collegato di Salerno, I-84084 Fisciano (SA), Italy}
\affiliation{$^3$ Dipartimento di Ingegneria Industriale, Universit\`a degli Studi di Salerno, Via Giovanni Paolo II 132, I-84084 Fisciano (SA), Italy}
\affiliation{$^4$ Department of Physics, University of Strathclyde, John Anderson Building, 107 Rottenrow, Glasgow G4 0NG, United Kingdom}

\date{July 28, 2015}

\begin{abstract}
We introduce a necessary and sufficient criterion for the non-Markovianity of Gaussian quantum dynamical maps based on the violation of divisibility.
The criterion is derived by defining a general vectorial representation of the covariance matrix which is then exploited to determine the condition for
the complete positivity of partial maps associated to arbitrary time intervals. Such construction does not rely on the Choi-Jamiolkowski representation and does not require optimization over states.
\end{abstract}


\maketitle


In recent years much effort has been devoted to the characterization and quantification of non-Markovianity in the evolution of open quantum systems
(see e.g. Ref.~\cite{RivasReview} for a recent review). Non-Markovian quantum evolutions may typically arise in the presence of structured
environments, such as in quantum biological systems~\cite{Bio1,Bio2,Bio3} and in squeezed baths of light with finite bandwidth~\cite{Zippilli2014}.
Moreover, recent studies suggest that properly engineered non-Markovian channels can improve the efficiency of quantum technology protocols due
to the backflow of information from the environment to the system~\cite{NMAdvantagea,NMAdvantageb,NMAdvantagec,StrEnvironmenta,StrEnvironmentb,StrEnvironmentc,Haseli}.
Establishing whether noisy quantum evolutions are non-Markovian and therefore preserve some memory on the story of the system is of capital importance in
the field of quantum cryptography~\cite{GCryptography}.

Various approaches to the characterization and quantification of quantum non-Markovianity have been introduced in recent
years~\cite{RivasReview,Addis,RelativEntropies2}.
Most of them are witnesses, and thus rely on sufficient, but not necessary, conditions~\cite{RivasReview}. They usually are based on the non-monotonic behavior of certain quantities in the presence of memory effects~\cite{TraceDistance,Entanglement,Volume,Fisher,Capacity,RelativEntropies}. Moreover, most of them rely on optimization over states.

Proper measures of non-Markovianity have also been introduced for finite-dimensional systems. These measures include the amount of isotropic noise necessary to make the dynamics completely positive in every arbitrary short interval of time~\cite{IsotropicNoise} and the negativity of the decay rates appearing in the generators of the time evolution, once the associated master equation is expressed in canonical form~\cite{Cresser2014}.

A further necessary and sufficient criterion has been obtained by Rivas, Huelga, and Plenio (RHP) by considering the violation of the divisibility property, which expresses the possibility of decomposing the evolution on a generic time interval into two successive, independent completely positive maps. Non-Markovianity is then characterized by the extent that the intermediate map violates complete positivity (CP)~\cite{Entanglement}. These three necessary and sufficient criteria for finite-dimensional systems have been shown to be completely equivalent~\cite{Cresser2014}; moreover, the
criterion based on the isotropic noise and the RHP criterion rely on positivity of the Choi-Jamio{\l}kowski states corresponding to the channels~\cite{Choi,Jamiolkowski}.

Addressing the general characterization and quantification of non-Markovianity in the infinite-dimensional case is an important open question, given the great importance of, for instance, Gaussian states and Gaussian channels in quantum optics, quantum information, and quantum technologies.
Inspired by the RHP approach, in the present work we introduce a necessary and sufficient criterion of non-Markovianity for Gaussian evolutions. It is
based on the violation of the divisibility property directly at the level of the matrices defining the channels, exploiting a powerful vectorial representation for them and for the covariance matrix. As such, the criterion does not require the use of the Choi-Jamio{\l}kowski isomorphism between states and channels~\cite{GaussianCJ}. Moreover, it does not require optimization over states, a challenging task both for finite- and infinite-dimensional systems~\cite{GFidelity}.


Based on such criterion, we introduce the corresponding measure of non-Markovianity for Gaussian channels and illustrate it for some paradigmatic examples. For the specific channels considered, violation of divisibility turns out to be equivalent to the negativity of the decoherence rates appearing in the canonical form of the master equation.

Given a generic input state, its time evolution in a Gaussian channel is defined according to the following transformation on the input covariance matrix $\sigma(0)$:
\begin{equation}\label{ChannelDef}
\sigma\left(t\right) = X\left(t\right) \sigma\left(0\right)X^\intercal\left(t\right)+Y\left(t\right)\; ,
\end{equation}
where $(X,Y)$ are $2N\times 2N$ real matrices; moreover $Y$ is symmetric. It is possible to show that the CP requirement imposes the condition~\cite{CPCond}:
\begin{equation}\label{XYCP}
Y\left(t\right)-\dfrac{\imath}{2}\Omega +\dfrac{\imath}{2}X\left(t\right)\Omega X^\intercal\left(t\right)\geq 0 \; ,
\end{equation}
where $\Omega$ is the symplectic matrix, and the symbol $\intercal$ denotes matrix transposition.

Gaussian channels enjoy a semigroup structure. Given two such channels corresponding, according to Eq.~(\ref{ChannelDef}), to the pairs $(X_1, Y_1)$ and $(X_2,Y_2)$, the resulting composed channel is characterized as follows~\cite{SemigroupStr}:
\begin{equation}\label{Concatenation}
(X_1,Y_1)\cdot(X_2,Y_2)=(X_1 X_2,X_1 Y_2X_1^\intercal+Y_1) \; .
\end{equation}

Consider now a quantum evolution from time $t_0$ to $t_2$ described, in general, by some family of trace-preserving linear maps $\{ {\mathcal{E}}(t_2,t_1),t_2\geq t_1\geq t_0 \}$. The map is said to be divisible, or Markovian, if, for every $t_2$ and $t_1$ it holds that
\begin{equation}\label{divisibility}
{\mathcal{E}}(t_2,t_0) = {\mathcal{E}}(t_2,t_1){\mathcal{E}}(t_1,t_0),\qquad t_2\geq t_1\geq t_0 \; ,
\end{equation}
and the map ${\mathcal{E}}(t_2,t_1)$ is CP. An evolution is non-Markovian if it violates the divisibility property, Eq.~(\ref{divisibility}).

For Gaussian channels, we can reformulate the general divisibility condition, Eq.~(\ref{divisibility}), in the following form. Let us first introduce an auxiliary vectorial notation. In this notation the elements of the covariance matrix $\sigma$ form a vector according to a lexicographical ordering, i.e. $(\vec{\sigma}(t))_k=\sigma(t)_{ij}$ (where $k=N(i-1)+j$ and $i,j=1,\ldots ,2N$), which in the more familiar Dirac notation can be written as $\langle k|\vec{\sigma}(t)\rangle\equiv\langle ij|\vec{\sigma}(t)\rangle\equiv \langle i|\sigma(t)|j\rangle$. For convenience, we also add to $\vec{\sigma}$ an auxiliary vector entry of value $1$.
In this notation one can obtain the following representation:~$[X(t)\sigma(0)X^\intercal(t)]_{ij}\!\!=\!\! $~$ [(X(t)\otimes $ $ X(t))\vec{\sigma}(0)]_k\equiv[\Phi(t)\vec{\sigma}(0)]_k$, where $\Phi(t)=X(t)\otimes X(t)$. In Dirac notation, one has:
\begin{eqnarray*}
&&\!\!\!\!\!\!\!\!\!\langle i|X\!(t)\sigma(0)X^\intercal\!(t)|j\rangle\!=\!\sum_{n,m}\langle i|X\!(t)|n\rangle\langle n|\sigma\!(0)|m\rangle\langle m|X^\intercal\!(t)|j\rangle\\
&\!\!\!=&\!\!\!\!\sum_{n,m}\langle ij| X\!(t)\!\otimes\! X\!(t)|nm\rangle\langle nm|\vec{\sigma}\!(0)\rangle\!=\!\langle ij|X\!(t)\!\otimes\! X\!(t)|\vec{\sigma}\!(0)\rangle,
\end{eqnarray*}
where $\sum_n|n\rangle\langle n|$ is the identity resolution in some basis. One can now reexpress Eq.~(\ref{ChannelDef}) in terms of a vector by matrix multiplication:
\begin{equation}\label{vectoper}
\left(\begin{array}{c}
\vec{\sigma}(t) \\
1
\end{array} \right) = \left(\begin{array}{cc}
\Phi(t) & \vec{Y}(t) \\
\vec{0}^\intercal & 1
\end{array} \right)\left(\begin{array}{c}
\vec{\sigma}(0) \\
1
\end{array} \right)\; ,
\end{equation}
where $\vec{0}=(0,\ldots,0)^\intercal$ is the $2N$-dimensional null vector and $\vec{Y}(t)$ is the vectorial form of the matrix $Y(t)$. Vectorization is an isomorphism, thus reversible: de-vectorizing Eq.~(\ref{vectoper}) yields exactly the standard representation, Eq.~(\ref{ChannelDef}).

In vectorial notation, the channel composition law, Eq.~(\ref{Concatenation}), is reexpressed by the following matrix multiplication form:
\begin{equation}
\left(\begin{array}{cc}
\Phi_2 & \vec{Y}_2 \\
\vec{0}^\intercal & 1
\end{array} \right)
\left(\begin{array}{cc}
\Phi_1 & \vec{Y}_1 \\
\vec{0}^\intercal & 1
\end{array} \right)=
\left(\begin{array}{cc}
\Phi_2\Phi_1 & \Phi_2\vec{Y}_1+\vec{Y}_2 \\
\vec{0}^\intercal & 1
\end{array} \right).
\end{equation}

Setting for ease of notation $t_0=0$, $t_1=t$, and $t_2 = t+\epsilon$ for any instance of time $t$ and $\epsilon$, by the continuity of time the dynamics can be split as $[0,t+\epsilon] = [0,t]\cup[t,t+\epsilon]$, and one can obtain the vectorial expression for the intermediate map in the interval $[t,t+\epsilon]$. 
Let us comment on the invertibility of the $X(t,0)$ matrix. Examples of Gaussian channels characterized by a non invertible $X$ matrix can be found based on classification of one-mode Gaussian channels provided in Ref.~\cite{Holevo2007}. Up to Gaussian unitary equivalence, channels for which $X$ is non invertible include the completely depolarising channel which projects every input state on a thermal state, and channels which transform the canonical quadrature $Q$ and $P$ as: $P\rightarrow p$, $Q\rightarrow Q+q$, where $p$ and $q$ are thermal states. However, non invertible cases do not impose any restriction on our procedure, because one can always introduce the matrix $\mathbb{1}\eta + X(t,0)$, determine its inverse, and evaluate the limit $\eta \rightarrow 0$, which is always non-singular~\cite{RivasReview,Bhatia,HornJohnson}.

De-vectorizing the intermediate Gaussian map, we obtain its complete expression in terms of the $X$ and $Y$ matrices:


\begin{align}\label{MatricesEvolution}
&X\!\left(t\!+\epsilon,t\right)\!\!=\!\!X\!\left(t\!+\!\epsilon,0\right)\!X^{-1}\!\left(t,0\right), \nonumber\\
\\
&Y\!\left(t\!+\!\epsilon,t\right)\!\!=\!\!Y\!\left(t\!+\!\epsilon,0\right)\!\!-\!\!X\!\left(t\!
+\!\epsilon,t\right)\!Y\!\left(t,0\right)\!X^\intercal\!\left(t\!+\!\epsilon,t\right)\nonumber.
\end{align}

Since the condition of divisibility is equivalent to CP of the intermediate map, Eq.~(\ref{MatricesEvolution}), from Eqs.~(\ref{XYCP}), (\ref{divisibility}), and (\ref{MatricesEvolution}), the condition for non-Markovianity at any given time $t$ reads:
\begin{equation}\label{NMC}
Y\left(t+\epsilon,t\right)-\dfrac{\imath}{2}\Omega +\dfrac{\imath}{2}X\left(t+\epsilon,t\right)\Omega X^\intercal\left(t+\epsilon,t\right) < 0 \; .
\end{equation}
Given that Eq.~(\ref{XYCP}) is necessary and sufficient for CP~\cite{CPCond}, it follows that Eq.~(\ref{NMC}) is a necessary and sufficient criterion for the non-Markovianity of Gaussian channels.

The necessary and sufficient condition, Eq.~(\ref{NMC}), allows to introduce a proper measure of non-Markovianity for Gaussian channels by quantifying the extent by which the intermediate dynamics fails to be CP. This corresponds to the quantification of the negative part of the spectrum of the symmetric matrix appearing on the l.h.s. of Ineq.~(\ref{NMC}). Denoting the set of eigenvalues by $\{\nu_k(t+\epsilon,t)\}_{k=1,...,2N}$, the following functions
\begin{equation}\label{Smallf}
f_k\left(t\right)=\dfrac{1}{2}\lim_{\epsilon\rightarrow 0^+}\left[\vert\nu_k(t+\epsilon,t)\vert-\nu_k(t+\epsilon,t)\right]
\end{equation}
quantify the negative contribution at time $t$ given by the $k$th eigenvalue. Therefore punctual non-Markovianity, quantified by the negative part of the spectrum at a given time $t$, reads
\begin{equation}\label{Bigf}
F\left(t\right)\equiv\sum_{k=1}^{2N}f_k\left(t\right)\; .
\end{equation}
Since $F(t)>0$ if and only if the evolution is non-Markovian, and $F(t) = 0$ otherwise, total non-Markovianity on a generic time interval $I$ is
\begin{equation}\label{Measure}
\mathcal{N}^I\equiv\int_I F\left(t\right)\;dt \; .
\end{equation}
It is important to note that, when the dynamics is described by means of a master equation, the expression of the matrices $\{X,Y\}$ that define the channel are obtained directly, in the phase space formalism, through Eq.~(\ref{ChannelDef}), from the expression of the characteristic function of the evolved Gaussian state.

The above divisibility-based necessary and sufficient criterion is completely general: it holds for any Gaussian map, independently of the existence of a generator. On the other hand, as already mentioned, Hall, Cresser, Li, and Andersson have recently shown that in the finite-dimensional case, for which, at variance with the infinite-dimensional case, all processes always admit a generator, the necessary and sufficient criterion for non-Markovianity based on divisibility is equivalent to the criterion based on the negativity of the decoherence rates appearing in the canonical form of the master equation~\cite{Cresser2014}. It is then tempting to conjecture that this equivalence holds also in the infinite-dimensional case for channels that admit a generator. 
we discuss two paradigmatic cases that admit a representation in terms of canonical master equations, pure damping and quantum Brownian motion, and show that in such instances the equivalence indeed holds.




The simplest example we can begin with is the Lindblad-type master equation describing the damping process for a single field mode with a single decay rate:
\begin{equation}
\dfrac{d\rho\left(t\right)}{dt}=\alpha\,\gamma\left(t\right)\left[a\rho a^\dag-\dfrac{1}{2}\left\lbrace a^\dag a, \rho\right\rbrace\right] \; ,
\end{equation}
where $\alpha\ll 1$ is the coupling constant and $\gamma(t)$ is the damping rate. The evolution of a generic Gaussian state in this Gaussian channel is described by the corresponding evolution of the displacement and covariance matrices. From the latter, via Eq.~(\ref{ChannelDef}), one obtains the $X$ and $Y$ matrices:
\begin{align}
X\left(t,0\right)&=e^{-\frac{\Gamma(t)}{2}}\mathbb{1}\; ,\label{DampX} \\
Y\left(t,0\right)&=\left[1-e^{-\Gamma\left(t\right)}\right]\dfrac{\mathbb{1}}{2} \; ,
\label{DampY}
\end{align}
where $\Gamma(t) = 2\alpha\int_0^t \gamma(s)ds$.
Eqs.~(\ref{DampX}), ~(\ref{DampY}) allow to obtain, through Eqs.~(\ref{MatricesEvolution}), the matrix appearing in the l.h.s. of the CP condition, Ineq.~(\ref{NMC}). It is straightforward to verify that the eigenvalues of this matrix are negative when $\exp(-\Gamma(t+\epsilon,t))< 1$, where $\Gamma(t+\epsilon,t) = \Gamma(t+\epsilon,0)-\Gamma(t,0)$. Moreover, to first order in $\epsilon$, we have $\Gamma(t+\epsilon,t)\approx 2\gamma(t)\epsilon$. As a consequence, the evolution is non-Markovian if and only if $\gamma(t)<0$, showing that in this case violation of divisibility is indeed equivalent to negativity of the decoherence rate. The corresponding measure, through Eqs.~(\ref{Smallf}) and~(\ref{Bigf}) reads:
\begin{equation}
\mathcal{N}^I=-\alpha\int_{I'}\gamma\left(t\right)dt \; ,
\label{DampingN}
\end{equation}
where $I'$ are the sub-intervals of $I$ in which $\gamma(t)<0$.


We next consider Quantum Brownian Motion in the weak coupling limit and under the secular approximation. It is described in the interaction picture by the following Lindblad-type master equation (see Ref.~\cite{QBMSolution} and references therein):
\begin{multline}\label{QBMEq}
\dfrac{d\rho\left(t\right)}{dt}=\dfrac{\Delta\left(t\right)+\gamma\left(t\right)}{2}\left[2 a\rho a^\dag-\left\lbrace a^\dag a,\rho\right\rbrace\right]+\\
\dfrac{\Delta\left(t\right)-\gamma\left(t\right)}{2}\left[2 a^\dag\rho a-\left\lbrace a a^\dag,\rho\right\rbrace\right] \; .
\end{multline}
The coefficients $\gamma(t)$ and $\Delta(t)$, in general time-dependent, are respectively the damping and diffusion coefficient, whose explicit expressions are obtained once one selects the explicit form for the spectral density of the bath.
The general solution of Eq.~(\ref{QBMEq}) allows to obtain the evolution of the displacement and covariance matrices for any input Gaussian state~\cite{QBMSolution}. The corresponding $X$ and $Y$ matrices read:
\begin{align}
X\left(t,0\right)&=e^{-\frac{\Gamma(t)}{2}}R\left(t\right) \; , \\
Y\left(t,0\right)&=e^{-\Gamma(t)}\tilde{\Delta}\left(t\right)\mathbb{1} \; ,
\end{align}
where $\Gamma(t)=2\int_0^t\gamma(s)ds$, $\tilde{\Delta}(t)=\int_0^te^{\Gamma(s)}\Delta(s)ds$, $R(t)$ is the rotation matrix by the angle $\omega_0 t$, and $\omega_0$ is the system's characteristic frequency.
These expressions and Eqs.~(\ref{MatricesEvolution}) determine the eigenvalues of the matrix in the l.h.s. of Ineq.~(\ref{NMC}):
\begin{align}\label{QBMEigenvalues}
\nu_1(t+\epsilon,t)&=\dfrac{1}{2}\left[e^{-\Gamma(t+\epsilon,t)}+2\tilde{\Delta}\left(t+\epsilon,t\right)e^{-\Gamma(t+\epsilon,0)}-1\right], \nonumber\\
\nu_2(t+\epsilon,t)&=\dfrac{1}{2}\left[1-e^{-\Gamma(t+\epsilon,t)}+2\tilde{\Delta}\left(t+\epsilon,t\right)e^{-\Gamma(t+\epsilon,0)}\right],
\end{align}
where $\Gamma(t+\epsilon,t)=\Gamma(t+\epsilon,0)-\Gamma(t,0)$ and $\tilde{\Delta}(t+\epsilon,t)=\tilde{\Delta}(t+\epsilon,0)-\tilde{\Delta}(t,0)$. To first order in $\epsilon$, we have: $e^{-\Gamma(t+\epsilon,t)}\approx 1-2\gamma(t)\epsilon$ and $\tilde{\Delta}(t+\epsilon,t)\approx e^{\Gamma(t,0)}\Delta(t)\epsilon$. Then, condition Eq.~(\ref{NMC}) on the eigenvalues, i.e. the violation of the divisibility condition, implies $\Delta(t)<|\gamma(t)|$. This is again equivalent to negativity of the decoherence rates $[\Delta(t)+\gamma(t)]/2$ and $[\Delta(t)-\gamma(t)]/2$ appearing in Eq.~(\ref{QBMEq}). Finally, exploiting Eqs.~(\ref{QBMEigenvalues}) and~(\ref{Smallf}) we obtain the following expression for the punctual measure of non-Markovianity:
\begin{equation}\label{QBMF}
F\left(t\right)=\dfrac{1}{2}\left[|\Delta\left(t\right)-\gamma\left(t\right)|+|\Delta\left(t\right)+\gamma\left(t\right)|\right]-\Delta\left(t\right) \; .
\end{equation}
In order to investigate explicitly the behavior of non Markovianity in the Quantum Brownian Motion, we need to specify the spectral density to obtain explicit expressions of the damping and diffusion coefficients $\gamma(t)$ and $\Delta(t)$. Considering the rather typical case of an Ohmic bath with an exponential cut-off $\omega_c$,
the parameters that govern the dynamics are the temperature $T$ and the ratio between the cut-off frequency of the bath and the characteristic frequency of the system $x = \omega_c / \omega_0$. It is expected that, in the regime $x\ll 1$ the dynamics should be non-Markovian, while Markovianity should be recovered for $x\gg 1$~\cite{QBMSolution}. It is also convenient to express the evolution in terms of the dimensionless reduced time $\tau=\omega_c t$. Moreover, explicit analytic expressions of the diffusion coefficient $\Delta(\tau)$ can be obtained quite straightforwardly in the high- and low-temperature regimes~\cite{Vasile2009}.
\begin{figure}[!ht]
\centering%
\subfigure%
{\includegraphics[scale=0.35]{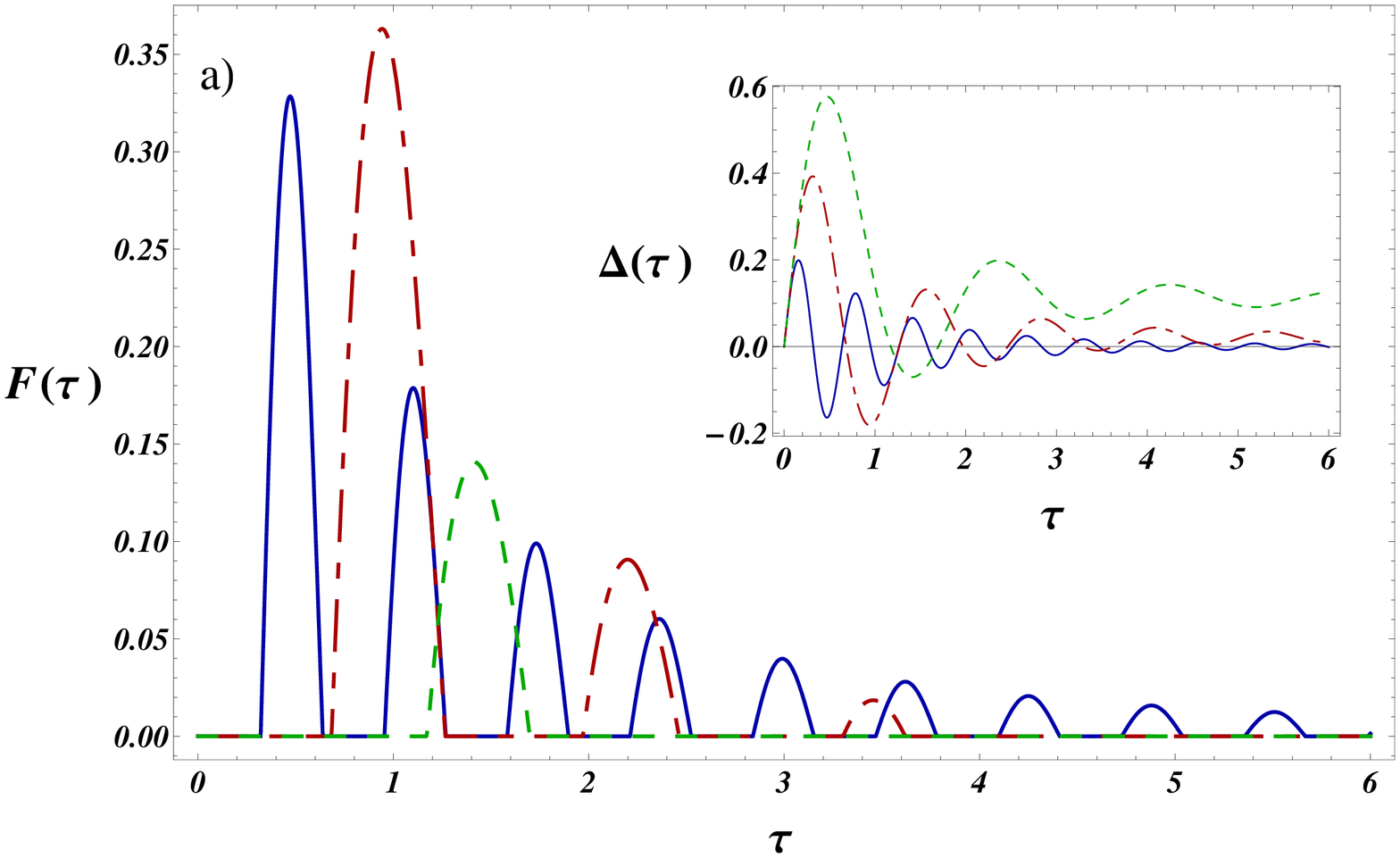}}
\subfigure%
{\includegraphics[scale=0.35]{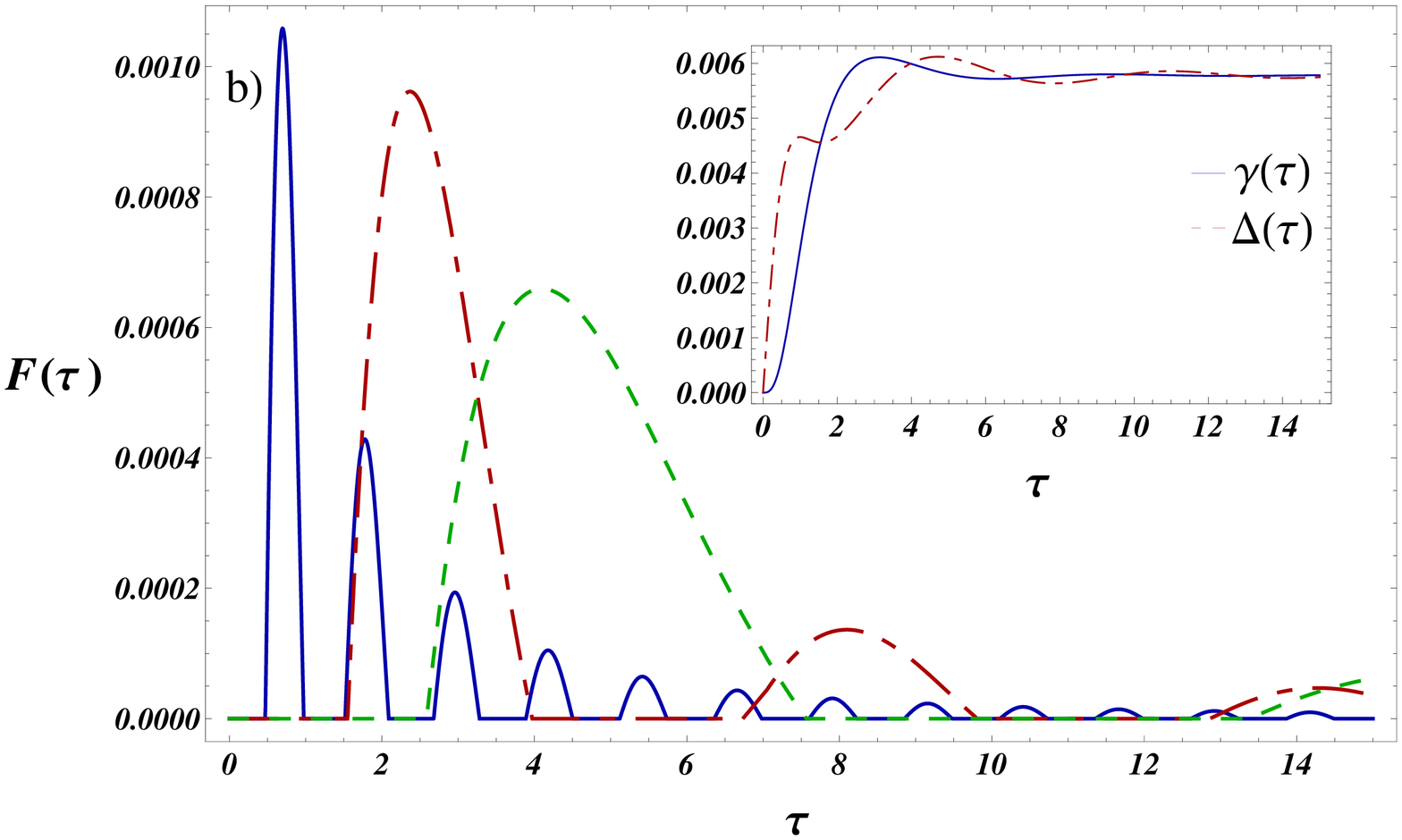}}
\caption{(color online) Non-Markovianity $F(\tau)$. $a)$: High-temperature limit for $x=0.1$ (blue full line), $x=0.2$ (red dot-dashed line), and $x=0.3$ (green dashed line). Inset: $\Delta(\tau)$ for $x=0.1$ (blue full line), $x=0.2$ (red dot-dashed line) and $x=0.3$ (green dashed line). $b)$: Low-temperature limit for $x=0.2$ (blue full line), $x=1.0$ (red dot-dashed line) and $x=2.0$ (green dashed line). Inset: $\Delta(\tau)$ (red dot-dashed line) and $\gamma(\tau)$ (blue line) for $x=0.1$.
\label{fig:QBMFT10}}
\end{figure}
Considering first the asymptotic values of the damping and diffusion coefficients in the large-time limit, $\tau \rightarrow \infty$, both in the high- and low-temperature regimes, it is straightforward to verify that
the asymptotic punctual non-Markovianity $F(\infty)=0$: at large times Markovianity is always recovered, independently of the values of the parameters that govern the dynamics.

Considering now generic times, in Fig.~(\ref{fig:QBMFT10}$a$) we report the behavior of the punctual non-Markovianity $F$, Eq.~(\ref{QBMF}), as a function of the reduced time $\tau$ at fixed values of the parameter $x=\omega_c/\omega_0$ in the high-temperature limit. In this regime $\Delta(\tau)\gg\gamma(\tau)$~\cite{QBMSolution}, and the non-Markovianity of the dynamics depends essentially only on the diffusion coefficient: $F(\tau)\simeq$ $|\Delta(\tau)|-\Delta(\tau)$. Hence, the time interval for which the evolution is non-Markovian, $F(\tau)>0$, corresponds to the negativity of the decoherence rate, $\Delta(\tau)<0$. Non-Markovianity is strongest in the regime $x\ll 1$, corresponding to the characteristic time of the bath being much larger than the characteristic time of the system. When $x$ increases, the negative part of the oscillations and $F(\tau)$ quickly vanishes, and one recovers the Markovian regime.

In the low-temperature regime, see Fig.~(\ref{fig:QBMFT10}$b$), the diffusion and damping coefficients are comparable, and the non-Markovianity $F(\tau)$ is given by the full expression, Eq.~(\ref{QBMF}). In this situation, a non-Markovian regime is observed also for $\Delta(\tau)> 0$, provided $\Delta(\tau) < \gamma(\tau)$, and even if the characteristic times of the bath start to be comparable or smaller than the characteristic times of the system, $x \gtrsim 1$. 




In these examples, the criterion based on the $X$ and $Y$ matrices defining a Gaussian channel turns out to correspond to the negativity of the decoherence rates. On the other hand, it should be stressed that the criterion is much more general and applies to any Gaussian evolution, including those that do not admit a generator and hence cannot be described in terms of master equations. Finally, it always allows, at least in principle the experimental verification of the Markovianity of the evolution. A further advantage is that such verification does not require optimization over the set of input states, since it is based directly on the characteristic matrices that define intrinsically the dynamical map. When the generator exists, so that the dynamical map can be associated to a master equation, the general quantifiers of non-Markovianity, Eqs.~(\ref{Smallf}) and~(\ref{Measure}), reduce to simple functions of the decoherence rates, Eqs.~(\ref{DampingN}) and~(\ref{QBMF}), which can be reconstructed experimentally~\cite{NMTomography}.

Summarizing, checking for CP of the partial map in the finite-dimensional case consists in verifying the positivity of the corresponding Choi-Jamio{\l}kowski state. The RHP measure of non-Markovianity is then defined in terms of the negative part of the spectrum of such state~\cite{RivasReview,Entanglement}. One might try to enforce the same criterion, in complete analogy with the finite-dimensional case, by checking CP of the intermediate Gaussian map by checking the positivity of the corresponding Choi-Jamio{\l}kowski state. Indeed, for single-mode CP Gaussian maps, it has been shown that a Kraus decomposition can always be found, so that one can always construct the corresponding Choi-Jamio{\l}kowski state~\cite{IvanKraus,GaussianCJ}. Unfortunately, the Kraus representation does not exists for non CP Gaussian maps and the Choi-Jamio{\l}kowski states corresponding to these maps are to date not characterized. Therefore, checking violation of CP for Gaussian maps using the Choi-Jamio{\l}kowski isomorphism is currently impossible.

We have succeeded in circumventing this stumbling block by expressing the condition of CP of general Gaussian maps, Ineq.~(\ref{XYCP}), directly in terms of the $X$ and $Y$ matrices governing the evolution of the covariance matrix of input Gaussian states.
Concatenation of the maps in the finite-dimensional case is expressed straightforwardly by matrix multiplication, thanks to the superoperator representation of the quantum channel. Immediate generalization of this method to general Gaussian maps is not possible. Instead, we succeed in defining Gaussian state vectorization by a suitable vectorization of the covariance matrix. This allows to introduce a matrix representation also for Gaussian maps and generalize the RHP method of characterizing non-Markovianity.

The specific form of the vectorization procedure that we have introduced is suitably defined in such a way to preserve the fundamental semigroup property of Gaussian channels, allowing to investigate any Gaussian map in a compact and elegant form.

Since our approach does not require optimization on the set of input states, it can be especially helpful when considering multi-mode Gaussian channels, allowing in principle for a systematic study of the interplay between non-Markovianity, entanglement, and coherence. Furthermore, it can open the way to the characterization of Gaussian quantum metrology~\cite{GaussIntPow} in non-Markovian environments, extending to Gaussian states of continuous-variable systems the investigations pioneered in Refs.~\cite{Matsuzaki,NonMarkMetr}.

{\em Acknowledgments --} We would like to thank Ra\'ul Garc\'ia-Patr\'on S\'anchez for useful discussions. We acknowledge the EU FP7 Cooperation STREP Projects iQIT - integrated Quantum Information Technologies, Grant Agreement No. 270843, and EQuaM - Emulators of Quantum Frustrated Magnetism, Grant Agreement No. 323714. We also acknowledge financial support from the Italian Minister of Scientific Research (MIUR) under the national PRIN programme.


\end{document}